\documentclass[prd,aps,noeprint,twocolumn,amsmath,amssymb,floatfix,nofootinbib]{revtex4-1}

\usepackage{graphicx}
\usepackage{dcolumn}
\usepackage{bm}
\usepackage{latexsym}
\usepackage{color}
\usepackage{cancel}
\usepackage[colorlinks=true,linkcolor=blue,citecolor=blue]{hyperref}

\begin{document}

\title{Generalized Fermi acceleration}

\author{Martin Lemoine} \affiliation{Institut d'Astrophysique de Paris,\\
CNRS -- Sorbonne Universit\'e, \\
98 bis boulevard Arago, F-75014 Paris, France}

\date{\today}

\begin{abstract}  
In highly conducting astrophysical plasmas, charged particles are generically accelerated through Fermi-type processes involving repeated interactions with moving magnetized scattering centers. The present paper proposes a generalized description of these acceleration processes, by following the momentum of the particle through a continuous sequence of accelerated frames, defined in such a way that the electric field vanishes at each point along the particle trajectory. In each locally inertial frame, the Lorentz force affects the direction of motion of the particle, but the energy changes solely as a result of inertial corrections. This unified description of Fermi acceleration applies equally well in sub- and ultrarelativistic settings, in Cartesian or non-Cartesian geometries, flat or nonflat space-time. Known results are recovered in a variety of regimes -- shock, turbulent and shear acceleration -- and new results are derived in lieu of applications, {\it e.g.} nonresonant acceleration in relativistic turbulence, stochastic unipolar inductive acceleration and centrifugo-shear acceleration close to the horizon of a  black hole.
\end{abstract}

\maketitle

\section{Introduction}\label{sec:introd}
In a landmark paper of 1949, E. Fermi laid down the basic principles of particle acceleration in astrophysical plasmas~\cite{1949PhRv...75.1169F}, which have ever since served as a guide in high-energy astrophysics. Noting that the high conductivity of these plasmas implies a vanishing electric field in the reference frame of the plasma, E. Fermi argued that acceleration had to result from the nonuniform stirring motion of a magnetized medium. In this picture, the acceleration can be described in kinematical terms as a collision between a light fast-moving particle and a heavy slowly moving scattering center, the magnetic field playing the role of the collision agent.

The scattering of particles in the magnetized turbulence on both sides of a shock front where, by definition, the plasma velocity changes abruptly, represents a rather vivid illustration of this process~\cite{1983RPPh...46..973D,*1987PhR...154....1B,*Marcowith_2016}, and its realization at supernova remnant shock waves actually provides the likely origin of the bulk of cosmic rays~\cite{1977ICRC...11..132A,1977DoSSR.234.1306K,1978ApJ...221L..29B,1978MNRAS.182..147B, *1978MNRAS.182..443B}. There are of course a large variety of ways in which a particle can draw energy from a nonuniformly moving highly conducting plasma. The stochastic acceleration of particles in a turbulent magnetized fluid, for instance, is none other than the generalization of the original Fermi process to particles interacting with random waves or structures, {\it e.g.}~\cite{1966PhFl....9.2377K,*1966PhRv..141..186S,*1967PhFl...10.2620H,1991A&A...250..266A,1993PhyU...36.1020B} and references therein. 

More generally, the Fermi scheme can apply to any flow for which the two electromagnetic $4-$scalars verify $\mathbf{E}\cdot\mathbf{B}\,=\,0$ and $\mathbf{E}^2-\mathbf{B}^2\,<\,0$, because under such conditions, one can always boost to a frame in which $\mathbf{E}$ vanishes. This frame moves at velocity $\boldsymbol{\beta}_{\mathbf{B}}\,=\mathbf{E}\times\mathbf{B}/B^2$ and, unless ideal magnetohydrodynamics (MHD) applies, it does not necessarily coincide with the bulk velocity of the plasma. This allows one to extend the notion of Fermi acceleration to a variety of interaction processes between particles and electromagnetic fields. In reconnection configurations, for instance, particles can be accelerated at a fast rate along the parallel electric field in the diffusion region but they can also be accelerated through their Fermi-type interactions with moving magnetized structures in the dynamical outflows of those regions. As a matter of fact, in-depth studies indicate that a substantial fraction, if not most of the energization occurs through the latter processes in reconnecting flows, {\it e.g.}~\cite{2006Natur.443..553D,*2012SSRv..173..521H,*2019arXiv190108308G}.

Various methods have been employed to characterize the physics and the efficiency of acceleration processes in the test-particle limit. Some model the trajectory of individual particles in the electromagnetic environment, using an approximate transport equation, {\it e.g.}~\cite{1963RvGSP...1..283N},  an effective random walk, {\it e.g.}~\cite{1978MNRAS.182..147B, *1978MNRAS.182..443B,1981MNRAS.196..135P,2006ApJ...652.1044R}, or a quasilinear (Born type) picture, see~\cite{1966PhFl....9.2377K,*1966PhRv..141..186S,*1967PhFl...10.2620H} as well as ~\cite{1991A&A...250..266A,2002cra..book.....S,*2011ApJ...732...96S} and references therein. Some others derive effective transport equations from the more fundamental Boltzmann equation through a perturbative expansion in powers of the magnitude of distortions of the distribution function, {\it e.g.}~\cite{1967PhFl...10.2620H,1968Icar....8...54D,1972ApJ...172..319J,1975MNRAS.172..557S, 1985ApJ...296..319W,*1989ApJ...340.1112W,1989ApJ...336..243S,1993PhyU...36.1020B, 1993ApJ...405L..79W} or more recently~\cite{2018MNRAS.479.1747A,*2018MNRAS.479.1771A}, to extract the transport coefficients. For the purpose of concrete applications, one is generally interested in the first two moments in momentum space, {\it i.e.}, the mean 4-momentum change $\left\langle \Delta p^\alpha/\Delta t\right\rangle$ and the diffusion tensor $\left\langle \Delta p^\alpha\Delta p^\beta/\Delta t\right\rangle$ ($\alpha,\,\beta\,=\,0,\ldots,4$). 

The present paper proposes an alternative description of Fermi acceleration that does not rely on a perturbative scheme. Rather, it follows the particle journey through a continuous sequence of local rest frames, defined in such a way that the local electric field vanishes at each point along the trajectory. In this picture, the particle suffers pitch angle scattering from the Lorentz force in this local frame, which induces spatial diffusion but no energy gain, while it experiences an inertial force associated to the change in velocity of the fluid, which provides energy gain (or loss). By definition, a particle traveling in a fully rectilinear way is decoupled from the turbulence and thus cannot gain energy; spatial diffusion and inertial forces are thus intertwined fundamentals of Fermi acceleration.  Thanks to an effective model describing the trajectory of the particle in phase space, as a random walk in configuration space and as the evolution of the momentum through the local frames in which it suffers the inertial forces, it is possible to derive the total energy gain (or diffusion rate). In short, this description generalizes to a continuous flow the original description of Fermi acceleration as a sequence of discrete interactions with moving magnetic scattering centers. 

The present approach borrows tools from general relativity to characterize the locally inertial frames and their evolution in space-time. Although it makes the algebra somewhat cumbersome, it allows to deal with flows of complex velocity patterns, possibly beyond the reach of quasilinear theory, or with flows in complex geometries, in both sub- and ultrarelativistic limits. Explicit examples and new applications are provided in the following. The discussion is laid out as follows. Section~\ref{sec:genf} introduces the general formalism, which is then applied to turbulent flows with a random pattern of velocities in Sec.~\ref{sec:nres}, to flows with a nontrivial mean velocity structure in Sec.~\ref{sec:nonz}, and to flows in nontrivial geometries in Sec.~\ref{sec:geom}. Conclusions are provided in Sec.~\ref{sec:conc}. Units are such that $c=1$ and the metric signature $(-,+,+,+)$. Unless otherwise stated, the accelerated particles are assumed ultrarelativistic.

\section{General formalism}\label{sec:genf}

\subsection{Locally inertial frames}
In general relativistic kinetics, it proves convenient to use separate frames for configuration space variables $\mathsf{x}$ and for momenta $\mathsf{p}$, see {\it e.g.}~\cite{1985ApJ...296..319W,*1989ApJ...340.1112W,2013PhRvD..88b3011C}. Coordinates are described in a coordinate basis characterized by its metric $g_{\mu\nu}({\mathsf x})$, which may be nontrivial, either because of the use of curvilinear coordinates in flat space-time, or because of space-time curvature, or both. This coordinate basis is defined in general terms by a set of four-vectors $\left\{\mathsf{e}_{\mu}\right\}$, the index $\mu$ labeling the four-vector. Correspondingly, a $4-$displacement is written as: $\mathsf{d x}\,=\,\mathsf{e}_{\mu}{\rm d}x^\mu$, and the scalar product is expressed as: $\mathsf{e}_{\mu}\cdot \mathsf{e}_{\nu}\,=\,g_{\mu\nu}$, guaranteeing that ${\rm d}s^2\,=\,g_{\mu\nu}{\rm d}x^\mu{\rm d}x^\nu$. In the following, this reference frame will correspond to the lab frame and be labeled $\mathcal R_{\rm L}$.

In contrast, momenta are more conveniently described in locally inertial frames, that can be set up at any point of the manifold through orthonormal tetrads. These are characterized by sets of orthonormal four-vectors $\left\{\mathsf{e}_{\hat a}({\mathsf x})\right\}$, {\it i.e.}, $\mathsf{e}_{\hat a}({\mathsf x})\cdot\mathsf{e}_{\hat b}({\mathsf x})\,=\,\eta_{\hat a\hat b}$, with $\eta_{\hat a\hat b}$ the Minkowski metric. All throughout, hatted indices $\hat a,\,\hat b,\,\hat c\ldots$ are used to denote indices in this orthonormal frame, while nonhatted indices run over the coordinates in the lab frame $\mathcal R_{\rm L}$. One clear advantage of using locally inertial frames is to treat the physics of momentum as in flat space, up to the space-time dependence of the tetrad.

The connection between these frames is achieved by relating the basis vectors of one frame to the other:  $\mathsf{e}_{\mu}\,=\,{e^{\hat a}}_\mu\,\mathsf{e}_{\hat a}$, thereby defining the vierbein (tetrad components) ${e^{\hat a}}_\mu({\mathsf x})$.  Consequently, momenta in the lab frame and in the locally inertial frame are related to one another through
\begin{equation}
\hat p^{\hat a}\,=\,{e^{\hat a}}_\mu({\mathsf x})\,p^\mu,\quad\quad
p^\mu\,=\,{e^\mu}_{\hat a}({\mathsf x})\,\hat p^{\hat a}\,,
\label{eq:ptet}
\end{equation}
where ${e^\mu}_{\hat a}({\mathsf x})$ represents the inverse vierbein:
\begin{equation}
{e^{\hat a}}_\mu({\mathsf x})\,{e^\mu}_{\hat b}({\mathsf x})\,=\,{\delta^{\hat a}}_{\hat b}\,,\quad\quad
{e^\mu}_{\hat b}({\mathsf x})\,{e^{\hat b}}_\nu({\mathsf x})\,=\,{\delta^\mu}_\nu\,.
\label{eq:invtet}
\end{equation}
From now on, tensors in the locally inertial frame are written with a hat symbol to distinguish them from their counterparts in the lab frame.

For the problem at hand, we need to define two locally inertial frames: one that is comoving with the flow\footnote{More generally, it is understood here the frame in which the electric field locally vanishes. For the sake of simplicity, it is assumed here that  ideal MHD is a good approximation, hence this frame coincides with the local frame of rest.}, written $\widehat{\mathcal R}_{\mathsf u}$, and one that is defined at fixed coordinates in the lab frame, that we label $\overline{\mathcal R}_{\rm L}$.
The role of this latter is to set-up a locally inertial frame at zero velocity, in case the metric in the $\mathcal R_{\rm L}$ frame is not Minkowskian. In the coordinate basis $\mathcal R_{\rm L}$, the plasma $4-$velocity is written: $u^\mu\,=\,(\gamma_u,\mathbf{u})$; $\boldsymbol{\beta}_{\mathbf u}\,=\,\mathbf{u}/\gamma_u$ represents its $3-$velocity.  Quantities (and indices) defined in $\overline{\mathcal R}_{\rm L}$ are indicated with a bar symbol. The vierbein that connects quantities in $\overline{\mathcal R}_{\rm L}$ to quantities in ${\mathcal R}_{\rm L}$ is written ${{e_{\rm L}\,}^{\overline a}}_\mu$, and its inverse, ${{e_{\rm L}\,}^\mu}_{\overline a}$. Consequently, the four-velocity of the flow in this locally inertial frame is written: $\overline u^{\overline a}\,=\,{{e_{\rm L}\,}^{\overline a}}_\mu\,u^\mu$.  Given that $\widehat{\mathcal R}_{\mathsf u}$ and $\overline{\mathcal R}_{\rm L}$ are both locally inertial frames set up at the same space-time point, one can transform tensors from the former to the latter through a special relativistic Lorentz transform ${\Lambda^{\overline a}}_{\hat b}$: 
\begin{align}
{\Lambda^{\overline 0}}_{\hat 0}({\mathsf x})&\,=\,\overline u^{\overline 0}({\mathsf x})\,,\quad & {\Lambda^{\overline 0}}_{\hat \i}({\mathsf x})&\,=\,\overline u_{\hat \i}({\mathsf x})\,,\nonumber\\
{\Lambda^{\overline \i}}_{\hat 0}({\mathsf x})&\,=\,\overline u^{\overline \i}({\mathsf x})\,,\quad & 
{\Lambda^{\overline \i}}_{\hat \j}({\mathsf x})&\,=\,{\delta^{\overline \i}}_{\hat \j}\,+\,\frac{\overline u^{\overline \i}({\mathsf x})\,\overline u_{\hat \j}({\mathsf x})}{1+\overline u^{\overline 0}({\mathsf x})}\,.
\label{eq:tet}
\end{align}
In the above equations, indices $\hat \i,\, \hat \j,\overline \i,\overline \j\,=\,1,\,2,\,3$. The inverse Lorentz transform ${\Lambda^{\hat a}}_{\overline b}$ follows immediately from the above by the substitution $\overline u^{\overline \i}\,\rightarrow\, -\,\overline u^{\overline \i}$. 

The vierbein ${e^{\hat a}}_\mu({\mathsf x})$ is eventually defined as
\begin{eqnarray}
{e^{\hat a}}_\mu({\mathsf x})&\,=\,&{\Lambda^{\hat a}}_{\overline b}({\mathsf x})\,\, {{e_{\rm L}\,}^{\overline b}}_\mu({\mathsf x})\,,\nonumber\\
{e^\mu}_{\hat a}({\mathsf x})&\,=\,&{{e_{\rm L}\,}^\mu}_{\overline b}({\mathsf x})\,\,{\Lambda^{\overline b}}_{\hat a}({\mathsf x})\,.
\label{eq:vb}
\end{eqnarray}
Of course, if the geometry is trivial in $\mathcal R_{\rm L}$, meaning $g_{\mu\nu}\,=\,\eta_{\mu\nu}$, then the vierbein to the locally inertial frame at fixed coordinates is itself trivial, {\it i.e.}, ${{e_{\rm L}\,}^{\overline b}}_\mu\,=\,{\rm diag}\left(1,1,1,1\right)$.

The space-time dependence of the vierbein is characterized by its connection $\widehat \Gamma^{\hat a}_{\hat b\hat c}$, also called Ricci rotation coefficient,
\begin{equation}
\widehat \Gamma^{\hat a}_{\hat b\hat c}\,=\,- {e^\beta}_{\hat b}\,{e^\gamma}_{\hat c}\,{e^{\hat a}}_{\beta;\,\gamma}\,,
\label{eq:conn1}
\end{equation}
where the semicolon represents as usual a covariant derivative with respect to the metric $g_{\mu\nu}$ in the coordinate basis. Explicitly, therefore,
\begin{equation}
\widehat \Gamma^{\hat a}_{\hat b\hat c}\,=\,- {e^\beta}_{\hat b}\,{e^\gamma}_{\hat c}\,{e^{\hat a}}_{\beta,\gamma}\,+\,{e^{\hat a}}_\alpha\,{e^\beta}_{\hat b}\,{e^\gamma}_{\hat c}\,\Gamma^\alpha_{\beta\gamma}\,,
\label{eq:conn2}
\end{equation}
in terms of the $g_{\mu\nu}$ metric Christoffel symbols $\Gamma^\alpha_{\beta\gamma}$. Note that the Ricci rotation coefficients $\widehat\Gamma^{\hat a}_{\hat b\hat c}$ are not symmetric in the lower two indices. The Appendix provides the connection coefficients to the locally inertial frame of a generic flow in Cartesian coordinates and in flat space-time.

\subsection{Particle kinetics}
In the comoving locally inertial frame $\widehat{\mathcal R}_{\mathsf u}$, the particle momentum evolves according to
\begin{equation}
\frac{{\rm d}\hat p^{\hat a}}{{\rm d}\tau}\,=\,\frac{q}{m}\,
{{\widehat F}^{\hat a}}\!_{\hat b}\,\hat p^{\hat b}\,-\,
\widehat\Gamma^{\hat a}_{\hat b\hat c}\,\frac{\hat p^{\hat b}\,\hat p^{\hat c}}{m}\,,
\label{eq:dyn}
\end{equation}
where ${{\widehat F}^{\hat a}}{}_{\hat b}$ represents the electromagnetic tensor in the comoving frame and $\tau$ represents proper time. It is defined according to: ${\rm d}\tau\,=\,\left[-g_{\mu\nu}{\rm d}x^\mu{\rm d}x^\nu\right]^{1/2}\,=\,(m/p^t){\rm d}t$ in terms of coordinate time interval ${\rm d}t$ in $\mathcal R_{\rm L}$.

 By construction of the locally inertial comoving frame $\widehat{\mathcal R}_{\mathsf u}$, ${{\widehat F}^{\hat 0}}{}_{\hat b}\,=\,0$, hence the Lorentz force does not enter the equation for the time component of the momentum, which then evolves through the space-time dependent inertial corrections that derive from the space-time dependence of the velocity flow. One recovers, at the formal level, the need for a space-time-dependent flow to achieve acceleration.

The general idea of the proposed method is to follow the evolution of this time component of the momentum in the locally inertial comoving frame, all along the particle trajectory, which is described in configuration space in the coordinate basis. In the locally inertial frame, the Lorentz force only provides angular scattering and all energy gains or losses are captured by the inertial correction proportional to $\widehat\Gamma^{\hat a}_{\hat b\hat c}$. Provided the velocity structure of the flow, or its statistical moments in the case of turbulence, is known, one can integrate this energy gain or loss along the particle trajectory to calculate, {\it e.g.} the mean energy gain or loss and/or the diffusion coefficient.

 In the limit of weak perturbations, the trajectory of the particle could be approximated with the zeroth-order unperturbed trajectory, as in quasilinear calculations. The following discussion adopts a more general approach and describes the trajectory in configuration space as a succession of scattering events, with mean scattering time $t_{\rm s}$, and then integrates over a duration much larger than $t_{\rm s}$. This random walk allows one to go beyond the regime of validity of quasilinear theory at the price, however, of deriving results that depend directly on $t_{\rm s}$. 

Several remarks are in order here. Firstly, the inertial correction in Eq.~(\ref{eq:dyn}) depends on the square of the particle momentum; this is directly related to the choice of proper time as an affine parameter along the trajectory. In the lab frame, $p^t\,=\,m{\rm d}t/{\rm d\tau}$, so that, written in terms of lab frame time, the rhs would depend on only one power of the momentum. Secondly, ambiguities may arise in relativistic flows, regarding the frame in which one defines the scattering time, or even the frame in which one models the diffusion in momentum space. This will be addressed in specific cases further below. Finally, one should stress that the above Eq.~(\ref{eq:dyn}) incorporates all relevant inertial corrections and can be applied equally well in flat or curved space-time, in the sub- or ultrarelativistic flow velocity limit. 

Over a proper time interval $\Delta\tau$, the energy gain of a particle in the lab frame can thus be written
\begin{align}
\Delta p^t(\Delta\tau)\,&=\,\left[ {e^t}_{\hat a}(\Delta\tau)-{e^t}_{\hat a}(0)\right]\hat p^{\hat a}(0)\nonumber\\
&\quad\,\,+\,{e^t}_{\hat a}(\Delta\tau)\,\int_0^{\Delta\tau}\,{\rm d}\tau_1\,\,\frac{{\rm d}\hat p^{\hat a}}{{\rm d}\tau_1}\,.
\label{eq:dpt}
\end{align}
The first term on the rhs characterizes the first-order Fermi energy gain (or loss) associated to the change of frame between the initial and the final states: it vanishes if the frames at initial and final times coincide and in any case, remains bounded in time. The second term follows the history of inertial corrections that the particle undergoes along its trajectory and represents the main term of interest here. 

The mean rate of energy gain can thus be obtained as
\begin{align}
\left\langle\frac{\Delta p^t}{\Delta t}\right\rangle &\,=\,
\underset{\Delta t\,\rightarrow\,+\infty}{\rm lim}\,
\frac{1}{\Delta t}\,\left\langle{{e^t}_{\hat a}}(\Delta\tau)\,
\int_0^{\Delta\tau}\,{\rm d}\tau_1\,\,
\frac{{\rm d}\hat p^{\hat a}}{{\rm d}\tau_1}\right\rangle\,,
\label{eq:dpt2}
\end{align}
and how the average is calculated more explicitly depends on the assumptions that one makes on the velocity field and on the trajectory of the particle. Definite examples will be provided further below.

In a similar way, the second-order moment can be written
\begin{align}
\left\langle\Delta p^t\,\Delta p^t\right\rangle&\,=\, 
\left\langle{e^t}_{\hat a}(\Delta\tau){e^t}_{\hat b}(\Delta\tau)
\int_{0}^{\Delta\tau}{\rm d}\tau_1{\rm d}\tau_2\,
 \frac{{\rm d}{\hat p}^{\hat a}}{{\rm d}\tau_1}\frac{{\rm d}{\hat p}^{\hat b}}{{\rm d}\tau_2}\right\rangle \nonumber\\
&\quad\quad\quad\,+\,\ldots
\label{eq:diff1}
\end{align}
and the diffusion coefficient can be calculated accordingly, as explained further below. The unspecified terms of the above equation can be easily recovered from Eq.~(\ref{eq:dpt}); they do not contribute to diffusion.

\subsection{An example: Shock acceleration}
For the sake of illustration, and to make contact with a well-known case, consider the problem of Fermi-1 acceleration at a shock front. It proves convenient here to use the downstream frame, {\it i.e} the shocked plasma rest frame, as the laboratory frame. The shock front moves at velocity $-\boldsymbol{\beta_2}$ with respect to that frame toward the $+\boldsymbol{x}$ direction ($\beta_2\,<\,0$ corresponds to the $3-$velocity of downstream with respect to the shock). The upstream medium, {\it viz.} the unshocked plasma, moves towards $-\boldsymbol{x}$ at $3-$velocity $\boldsymbol{\beta_{\rm rel}}$; $\beta_{\rm rel}\,<\,0$ denotes the relative velocity between upstream and downstream. In terms of $\beta_1$, the velocity of upstream with respect to the shock front: $\beta_{\rm rel}\,=\,(\beta_1-\beta_2)/(1-\beta_1\beta_2)$.

In the above description, the $4-$velocity flow can be described as: $u^\mu\,=\,\left[u^t(x_{\rm s}),u^x(x_{\rm s}),0,0\right]$, with $u^x(x_{\rm s})\,=\,u_{\rm rel}\Theta(x_{\rm s})$, $x_{\rm s}=x+\beta_2t$ representing the distance to the shock, $u_{\rm rel}\,=\,\gamma_{\rm rel}\beta_{\rm rel}$ and $\Theta(x)$ the Heaviside function. To compute the mean acceleration rate, assume that the particle travels around the shock front, starting and ending its trajectory on the downstream side. In this case, the first term on the rhs of Eq.~(\ref{eq:dpt}) vanishes; furthermore, the vierbein at initial and final times is trivial, because the locally inertial frame at these instants matches the lab frame. Then, Eq.~(\ref{eq:dpt}) gives the variation $\Delta {p_2}^t$ between initial and final times as
\begin{equation}
\Delta {p_2}^t\,=\,\int{\rm d}\tau\,\frac{{\rm d}{\hat p}^{\hat 0}}{\rm d\tau}\,,
\label{eq:sho1}
\end{equation}
where the subscript $_2$ indicates quantities evaluated in the downstream frame.
The nonzero connection terms of $\widehat\Gamma^{\hat 0}_{\hat b\hat c}$ are: $\widehat\Gamma^{\hat 0}_{\hat 1\hat 0}\,=\, {u^x}_{,t}+\beta_u\,{u^x}_{,x}$ and $\widehat\Gamma^{\hat 0}_{\hat 1\hat 1}\,=\,\beta_u {u^x}_{,t} + {u^x}_{,x}$. Hence, Eq.~(\ref{eq:sho1}) reduces to
\begin{eqnarray}
\Delta {p_2}^t&\,=\,&-\int{\rm d}\tau\,\frac{\hat p^{\hat 1}}{m}
\left[\left(\hat p^{\hat 0} + \beta_u \hat p^{\hat 1}\right){u^x}_{,t}\,+\,
\left(\beta_u \hat p^{\hat 0} + \hat p^{\hat 1}\right){u^x}_{,x}\right]\nonumber\\
&\,=\,& -\int_{\mathcal C}{\rm d}x_{\rm s}\,\frac{\hat p^{\hat 1}}{\gamma_u} 
\frac{{\rm d}u^x}{{\rm d}x_{\rm s}}\,.
\label{eq:sho2}
\end{eqnarray}
Here $\mathcal C$ symbolizes the trajectory of the particle back and forth across the shock and $x_{\rm s}=x(\tau) + \beta_2 t(\tau)$. The quantity $\hat p^{\hat 1}$ depends on this coordinate because it depends on the history of the particle. The second equation follows by noting that ${p_2}^t=\gamma_u(\hat p^{\hat 0} + \beta_u \hat p^{\hat 1})$, ${p_2}^x=\gamma_u(\beta_u \hat p^{\hat 0}+\hat p^{\hat 1})$  and ${\rm d}\tau=(m/{p_2}^t){\rm d}t=(m/{p_2}^x){\rm d}x$; recall that $x$ and $t$ are defined in the lab frame which coincides with downstream. The above formula illustrates clearly that the total amount of energy gain scales with the number of shock crossings, since this is the only place where ${\rm d}{u^x}/{\rm d}x_{\rm s}$ does not vanish.

Noting that one can alternatively use
\begin{equation}
\frac{{\rm d}\hat p^{\hat 0}}{{\rm d}x_{\rm s}}\,=\,-\frac{\hat p^{\hat 1}}{\gamma_u}\frac{{\rm d}u^x}{{\rm d}x_{\rm s}},
\label{eq:sho3}
\end{equation}
the integral is reduced, without much surprise, as
\begin{equation}
\Delta {p_2}^t\,=\,\sum_{i_\rightarrow}\left[{p_1}^t-{p_2}^t\right]_{i_\rightarrow}
+\sum_{i_\leftarrow}\left[{p_2}^t-{p_1}^t\right]_{i_\leftarrow}\,,
\label{eq:sho5}
\end{equation}
where the symbol $i_\rightarrow$ (respectively, $i_\leftarrow$) numbers the shock crossings from downstream to upstream (respectively, upstream to downstream). Consider an ensemble of Fermi cycles, so that $i_\rightarrow$ jumps come in equal number to $i_\leftarrow$ jumps; group these together, and divide the above by $\Delta t$ to obtain the mean acceleration rate $\Delta {p_2}^t/\Delta t$ in the limit $\Delta t\,\rightarrow\,+\infty$:
\begin{equation}
\left\langle \frac{\Delta {p_2}^t}{\Delta t}\right\rangle\,=\,
\frac{1}{t_{\rm d\vert 2}+t_{\rm u\vert 2}}\,\left(\langle{p_1}^t-{p_2}^t\rangle_{i_\rightarrow}-\langle{p_1}^t-{p_2}^t\rangle_{i_\leftarrow}\right)\,.
\label{eq:sho6}
\end{equation}
The fraction that appears on the rhs is ${\rm d}N_{\rm cyc}/{\rm d}t$, the rate of Fermi cycles that the particle completes, and it is expressed as the inverse of the total time it takes to complete one such cycle; $t_{\rm d\vert 2}$ (respectively, $t_{\rm u\vert 2}$) correspondingly represents the mean time spent downstream (respectively, upstream), as measured in the downstream rest frame (this precision being of importance in a relativistic setting). In subrelativistic diffusive shock acceleration, $t_{\rm d\vert 2}\,\simeq\,(4/3)t_{\rm s\vert 2}/\vert\beta_2\vert$ and $t_{\rm d\vert 1}\,\simeq\,(4/3)t_{\rm s\vert 1}/\vert\beta_1\vert$~\cite{1983RPPh...46..973D}. The above formula may also be applied, however, to shock-drift acceleration, in which case $t_{\rm d\vert 2}\,\simeq\,t_{\rm g\vert 2}$ ($t_{\rm g\vert 2}$ gyration time in the downstream) and similarly for $t_{\rm d\vert 1}$.

The second term in brackets in Eq.~(\ref{eq:sho6}) is reduced by writing ${p_2}^t\,=\,\gamma_{\rm rel}{p_1}^t+u_{\rm rel}{p_1}^x$,
\begin{equation}
\left\langle \frac{\Delta {p_2}^t}{\Delta t}\right\rangle\,=\,
\frac{1}{t_{\rm d\vert 2}+t_{\rm u\vert 2}}\,u_{\rm rel}\left[\left\langle{p_1}^x\right\rangle_{i_\leftarrow}-\left\langle{p_1}^x\right\rangle_{i_\rightarrow}\right]\,.
\label{eq:sho7}
\end{equation}
In the subrelativistic limit $\vert u_{\rm rel}\vert\,\ll\,1$, the angular average yields~\cite{1983RPPh...46..973D}
\begin{equation}
\left\langle {p_1}^x\right\rangle_{i_\rightarrow}\,\simeq\,\frac{2}{3}{p_1}^t\,,
\label{eq:sho8}
\end{equation}
and the statistics at return $i_\leftarrow$ provide an opposite $-(2/3){p_1}^t$, so that
\begin{equation}
\left\langle \frac{\Delta {p_2}^t}{\Delta t}\right\rangle\,=\,-\frac{4}{3}
\frac{u_{\rm rel}}{t_{\rm d\vert 2}+t_{\rm u\vert 2}}\,{p_2}^t\quad (\vert u_{\rm rel}\vert\,\ll\,1)\,,
\label{eq:sho9}
\end{equation}
which is the standard result (recall $u_{\rm rel}\,<\,0$ here).

In the ultrarelativistic limit, the particle population remains highly anisotropic in the upstream plasma as a consequence of the large shock velocity, $\beta_1\,\sim\,-1$, which restricts the precursor to a size much less than the scattering length of particles and thereby effectively forbids large angular scattering~\cite{1999MNRAS.305L...6G,*2001MNRAS.328..393A}. Consequently, $\left.{p_1}^x\right\vert_{i\leftarrow}-\left.{p_1}^x\right\vert_{i\rightarrow}\,\approx\,-\left.{p_1}^x\right\vert_{i\rightarrow}/\gamma_1^2\,\approx\,{\mathcal O}(1)\,{p_2}^t/\gamma_{\rm rel}$, and
\begin{equation}
\left\langle \frac{\Delta {p_2}^t}{\Delta t}\right\rangle\,\approx\,-
\frac{{\mathcal O}(1)}{t_{\rm d\vert 2}+t_{\rm u\vert 2}}\,\beta_{\rm rel}\,{p_2}^t\quad (\vert u_{\rm rel}\vert\,\gg\,1)\,,
\label{eq:sho10}
\end{equation}
with $\beta_{\rm rel}\,\simeq\,-1$. This effectively matches the acceleration rate in the ultrarelativistic limit, {\it e.g.}~\cite{2003ApJ...589L..73L}.
\bigskip

\section{Nonresonant turbulent acceleration}\label{sec:nres}
Particles can gain energy from a magnetized turbulence through their interactions with the random electric fields. This process is generally described by gyroresonant or Landau-resonant particle-wave interactions~\cite{1966PhFl....9.2377K}, but such resonances are generically washed out by anisotropy effects in modern turbulence theories~\cite{2000PhRvL..85.4656C} and the very description of turbulence as a bath of linear waves is itself questionable. Meanwhile, particles can gain energy through nonresonant processes in a generic random turbulence flow.
For instance, compression regions with finite (and negative) $\boldsymbol{\nabla}\cdot\mathbf{u}$ give rise to energy gain while decompression regions give rise to energy loss. At the microscopic level, the compression stirs the fluid and generates a net electric field in the lab frame that accelerates the particle. How the velocity pattern of the flow affects the particle momentum is entirely encoded in Eq.~(\ref{eq:dyn}).

Consider for instance the force felt in the locally inertial frame by an isotropic population of particles. It is given by the average of Eq.~(\ref{eq:dyn}), assuming that $\mathbf{\hat p}$ is random. This leads to
\begin{equation}
\left\langle\frac{{\rm d}{\hat p}^{\hat 0}}{{\rm d}\tau}\right\rangle_{\mathbf{\hat p}}\,=\,-\frac{1}{3}\frac{{\hat p}^2}{m}\, \partial_\alpha u^\alpha\,,
\label{eq:avf}
\end{equation}
which correctly reproduces the average force exerted on a fluid by compressible and time-dependent motions~\cite{1989ApJ...340.1112W}. In the subrelativistic limit, the effect of time-changing velocity fields are usually neglected because they are of higher order in the velocity, but such terms cannot be neglected in the relativistic limit. 

 In the subrelativistic limit, the diffusion coefficients characterizing nonresonant diffusion processes have been evaluated in Refs.~\cite{1983ICRC....9..313B,1988SvAL...14..255P} using the nonrelativistic diffusive cosmic-ray transport equation~\cite{1965P&SS...13....9P,1975MNRAS.172..557S} and further studied in various limits in {\it e.g.}, Refs.~\cite{1990A&A...236..519D,*2003ApJ...595..195W,*2004ApJ...603...23C, *2006ApJ...638..811C,*2010ApJ...713..475J,*2013ApJ...777..128L,2013ApJ...767L..16O}. For a discussion of the relativistic limit, see {\it e.g.}~\cite{1996ApJ...461L..37B,1999A&A...350..705P}. Here we use the above formalism to generalize this process to relativistic flows, providing at the same time a new and more exhaustive derivation of the results in the subrelativistic limit.

The turbulence is characterized as follows: the magnitude of the four-velocity is assumed constant in time and space, {\it viz.} ${u^t}_{,\alpha}\,=\,0$; however, the three-vector $u^i$ is a random variable with zero mean and correlation tensor:
\begin{equation}
\left\langle u^i({\mathsf x_1})\,u^j({\mathsf x_2})\right\rangle\,=\,u^2\,\mathcal C_u^{ij}\left({\mathsf x_1};\,{\mathsf x_2}\right)\,.
\label{eq:corru}
\end{equation}
In the following, isotropic correlation functions are used, {\it e.g.} 
$\mathcal C_u^{ij}\left({\mathsf x_1};\,{\mathsf x_2}\right)\,=\,\frac{1}{3}\eta^{ij}\mathcal C_u\left({\mathsf x_1};\,{\mathsf x_2}\right)$ for simplicity, but these results can be generalized to anisotropic configurations. The equal-time correlation function at a given position is trivial, {\it i.e.}, $\mathcal C_u\left({\mathsf x_1};\,{\mathsf x_1}\right)\,=\,1$.

Write $\gamma_u\,=\,u^t$ and $\beta_u\,=\,\sqrt{1-1/\gamma_u^2}$ the fixed magnitude of the flow Lorentz factor and of its $3-$velocity. In the present case, the tetrad components ${e^{\hat a}}_\mu$ that relate the lab frame to the locally inertial frame $\widehat{\mathcal R}_{\mathsf u}$, where $\mathbf{\hat u}\,=\,0$, also become random fields. The constancy of $\beta_u$ implies that all terms of the form $u^j u_{j,k}$ vanish identically, which simplifies substantially the connection provided in the Appendix.

In such a random field, the average force exerted on a particle vanishes, see Eq.~(\ref{eq:avf}). The second-order moment, given in Eq.~(\ref{eq:diff1}), is however not trivial. This second-order moment comprises two terms, one of which expresses a typical "first-order" Fermi squared energy change, due to the difference of frames. Defining 
\begin{equation}
\Delta_0^2\,\equiv\, \left\langle\left[{e^t}_{\hat a}(\Delta\tau)-{e^t}_{\hat a}(0)\right]\left[{e^t}_{\hat b}(\Delta\tau)-{e^t}_{\hat b}(0)\right]{\hat p}^{\hat a}(0){\hat p}^{\hat b}(0)\right\rangle\,,
\label{eq:diff1b}
\end{equation}
this term is evaluated as
\begin{eqnarray}
\Delta_0^2&\,\simeq\,&p(0)^2\begin{cases}\displaystyle{ \frac{4}{3}}\beta_u^2\, & \quad \left(\gamma_u^2\beta_u^2\,\ll\,1\right)\,,\\
2\gamma_u^4\, & \quad\left(\gamma_u^2\beta_u^2\,\gg\,1\right)\,.\end{cases}
\label{eq:rhs2}
\end{eqnarray}
This calculation takes proper account of the correlations between ${\hat p}^{\hat a}$ and ${e^t}_{\hat b}$, as follows: it is assumed that ${\mathsf p}$ is initially uncorrelated with ${\mathsf u}$, so that ${\hat p}^{\hat a}(0)\,=\,{e^{\hat a}}_\alpha(0)\,p^\alpha(0)$ exhibits a partial correlation with ${e^t}_{\hat b}(0)$ through the correlation of the velocity field, but that at the final time, this correlation has disappeared, because the momentum direction has itself become randomized.

In the following, the above term is neglected because it remains bounded in time; hence, it does not contribute to diffusion.  The term of interest is thus the first one on the rhs of Eq.~(\ref{eq:diff1}), which can be simplified as follows. One  first notes that the direction of the spatial part of $\hat p^{\hat a}$ undergoes a random walk with mean scattering time $\hat t_{\rm s}$ in the locally inertial frame. At the final time, formally $\Delta t\,\rightarrow\,+\infty$, one can assume that $\hat p^{\hat a}$ is uncorrelated with the direction of the velocity of the flow, and that $\hat p^{\hat \i}$ is itself a random variable, with vanishing mean but nonvanishing correlation function $\left\langle \hat p^{\hat \i}\hat p^{\hat j}\right\rangle$. This leads to
\begin{align}
\left\langle\frac{\Delta p^t\Delta p^t}{2\Delta t}\right\rangle\,=\,&
\underset{\Delta t\,\rightarrow\,+\infty}{\rm lim}\,\frac{1}{2\Delta t}\int{\rm d}\tau_1{\rm d}\tau_2\,\gamma_u^2\left(1+\frac{1}{3}\beta_u^2\right)\nonumber\\
&\quad\quad\quad\quad\quad\times
\left\langle\frac{{\rm d}{\hat p}^{\hat 0}}{{\rm d}\tau_1}\frac{{\rm d}{\hat p}^{\hat 0}}{{\rm d}\tau_2}\right\rangle\,.
\label{eq:diff0}
\end{align}
Interestingly, the $\widehat{\mathcal R}_{\mathsf u}$-Lorentz force does not enter this expression, so that the energy gain is effectively provided by the (random) inertial correction associated to the connection, or, in other words, to the (random) effective gravity field. The term $\beta_u^2/3$ comes from the product $\left\langle {e^t}_{\hat \i}{e^t}_{\hat \j}\Delta \hat p^{\hat \i}\Delta \hat p^{\hat \j}\right\rangle$, noting that ${e^t}_{\hat a}\,=\,-u_a$ and that $\left\langle\Delta \hat p^{\hat \i}\Delta \hat p^{\hat \j}\right\rangle \,=\,\frac{1}{3}\eta^{\hat \i\hat \j}
\left\langle\Delta \hat p^{\hat 0}\Delta \hat p^{\hat 0}\right\rangle$ due to the isotropization of the momentum direction.

The above average involves up to eight factors of $u^i$ components, two factors of derivatives ${u^i}_{,a}$ and four factors of $\hat p^{\hat a}$. This average can be reduced, assuming that $u^i$ behaves as a Gaussian random field, so that, {\it e.g.}
\begin{equation}
\left\langle u^i u^j u^k u^l\right\rangle\,=\,{\mathcal C_u}^{ij} {\mathcal C_u}^{kl} +
{\mathcal C_u}^{ik} {\mathcal C_u}^{jl}+{\mathcal C_u}^{il} {\mathcal C_u}^{jk}\,,
\label{eq:brk}
\end{equation}
where the space-time dependence has not been made explicit for the sake of compactness, and similarly for higher-order functions. Assuming further that the velocities do not correlate with their derivatives: $\left\langle {u^i}_{,\mu} u^j\right\rangle\,=\,0$, the velocity derivatives can decomposed into compressive, shear and vorticity components as 
\begin{equation}
{u^i}_{,j}\,=\,\frac{1}{3}\,{\delta^i}_j\, \theta\,+\,{\sigma^i}_j\,+\,{\omega^i}_j\,,
\label{eq:decomp}
\end{equation}
where $\theta\,=\,\boldsymbol{\nabla}\cdot\mathbf{u}$ denotes the spatial compression / expansion, ${\sigma^i}_j$ represents the traceless symmetric shear tensor and ${\omega^i}_j$ the traceless antisymmetric vorticity tensor:
\begin{equation}
\sigma_{ij}\,=\,u_{(i,j)}-\frac{1}{3}\,\eta_{ij}\,\theta\,,\quad
\omega_{ij}\,=\,u_{[i,j]}\,,
\label{eq:sigom}
\end{equation}
with $u_{(i,j)}\,=\,\left(u_{i,j}+u_{j,i}\right)/2$, $u_{[i,j]}\,=\,\left(u_{i,j}-u_{j,i}\right)/2$. Finally, the quantity $a_i\,=\,u_{i,t}$ is used to represent the acceleration part.

Note that this decomposition differs from the general four-dimensional Helmholtz decomposition of $u^\mu$:
\begin{equation}
u_{\alpha,\beta}\,=\,{\sigma^{\scriptscriptstyle{ \rm 4D}}}_{\alpha\beta} \,+\,{\omega^{\scriptscriptstyle{ \rm 4D}}}_{\alpha\beta}\,+\,\frac{1}{3}\theta^{\scriptscriptstyle{ \rm 4D}}{h^{\scriptscriptstyle{ \rm 4D}}}_{\alpha\beta} + {a^{\scriptscriptstyle{ \rm 4D}}}_{\alpha\beta}\,,
\label{eq:cong}
\end{equation}
with
\begin{align}
{h^{\scriptscriptstyle{ \rm 4D}}}_{\alpha\beta}&\,=\eta_{\alpha\beta} + u_\alpha u_\beta\,,\nonumber\\
{\sigma^{\scriptscriptstyle{ \rm 4D}}}_{\alpha\beta}&\,=\,u_{(\alpha,\beta)} + u_{(\alpha}u^\mu u_{\beta),\mu} - \frac{1}{3}\theta {h^{\scriptscriptstyle{ \rm 4D}}}_{\alpha\beta}\,,\nonumber\\
{\omega^{\scriptscriptstyle{ \rm 4D}}}_{\alpha\,\beta}&\,=\,u_{[\alpha,\beta]} - u_{[\alpha}u^\mu u_{\beta],\mu}\,,\nonumber\\
{a^{\scriptscriptstyle{ \rm 4D}}}_{\alpha\beta}&\,=\,-u_\beta u^\mu u_{\alpha,\mu}\,.
\label{eq:cong1}
\end{align}
The difference stems from the terms involving three products of $u^\mu$: they do not appear in our expressions because of the assumption of Gaussian statistics, which breaks down the correlation functions to a minimum number of $u^\mu$. The present decomposition matches the standard nonrelativistic version, but it remains valid in the fully relativistic regime.

Under the assumption that the expansion, shear, vorticity and acceleration contributions are uncorrelated, the correlation functions that emerge in the calculations are
\begin{align}
\left\langle u^i({\mathsf x_1}) u_j({\mathsf x_2})\right\rangle&\,=\,\frac{1}{3}{\delta^i}_j\, \gamma_u^2\beta_u^2\, \mathcal C_u\left({\mathsf x_1};\,{\mathsf x_2}\right)\,,\nonumber\\
\left\langle u^{i,j}({\mathsf x_1}) u_{i,j}({\mathsf x_2})\right\rangle&\,=\,\frac{1}{3}\left\langle\theta^2\right\rangle\,\mathcal C_\theta\left({\mathsf x_1};\,{\mathsf x_2}\right) \nonumber\\&\,\quad+ \left\langle\sigma^2\right\rangle\,\mathcal C_\sigma\left({\mathsf x_1};\,{\mathsf x_2}\right) + \left\langle\omega^2\right\rangle\,\mathcal C_\omega\left({\mathsf x_1};\,{\mathsf x_2}\right)\,,\nonumber\\
\left\langle u^{i,j}({\mathsf x_1}) u_{j,i}({\mathsf x_2})\right\rangle&\,=\,\frac{1}{3}\left\langle\theta^2\right\rangle\,\mathcal C_\theta\left({\mathsf x_1};\,{\mathsf x_2}\right) \nonumber\\&\,\quad+ \left\langle\sigma^2\right\rangle\,\mathcal C_\sigma\left({\mathsf x_1};\,{\mathsf x_2}\right) - \left\langle\omega^2\right\rangle\,\mathcal C_\omega\left({\mathsf x_1};\,{\mathsf x_2}\right)\,,\nonumber\\
\left\langle {u^i}_{,i}({\mathsf x_1}) {u^j}_{,j}({\mathsf x_2})\right\rangle&\,=\, \left\langle\theta^2\right\rangle\, \mathcal C_\theta\left({\mathsf x_1};\,{\mathsf x_2}\right)\,,\nonumber\\
\left\langle {u^i}_{,t}({\mathsf x_1}) {u^j}_{,t}({\mathsf x_2})\right\rangle&\,=\,\frac{1}{3}\delta^{i j}\,\left\langle a^2\right\rangle\,\mathcal C_a\left({\mathsf x_1};\,{\mathsf x_2}\right)\,,
\label{eq:dcorr}
\end{align}
where $\mathcal C_u$, $\mathcal C_\nabla$, $\mathcal C_\sigma$, $\mathcal C_\omega$ and $\mathcal C_a$, respectively represent the velocity, divergence, shear, vortical and acceleration two-point functions; $\sigma^2\,=\,\sigma_{ij}\sigma^{ij}$ and $\omega^2\,=\,\omega_{ij}\omega^{ij}$. All these correlation functions depend on space and time, and in the lab frame, it is assumed that
\begin{equation}
\mathcal C_I({\mathsf x_1};\,{\mathsf x_2})\,=\,\exp\left[-\frac{\pi}{4}\Omega_I^2\left({t_1}-{t_2}\right)^2-\frac{\pi}{4}K_I^2\left(\mathbf{x_1}-\mathbf{x_2}\right)^2\right]\,,
\label{eq:corr}
\end{equation}
for $I\,\in\,\left\{u,\,\nabla,\,\sigma,\,\omega,\,a\right\}$. Note that the exact form of the correlation function does not matter much. The normalization guarantees that $\Omega_I$ and $K_I$, respectively, correspond to the inverse correlation time and length of the turbulence. 

The integral also depends on the evolution of momenta, in particular on their correlation function, which is formally expressed as
\begin{equation}
\left\langle \hat p^{\hat \i}(t_1) \hat p^{\hat \j}(t_2) \right\rangle\,=\,\frac{1}{3}\eta^{\hat \i\hat \j}\,\hat p_1\hat p_2\,\mathcal C_p(t_1;\,t_2)\,,
\label{eq:Cp}
\end{equation}
where $\hat p_{1,2}\,=\,\hat p(\tau_{1,2})$, and 
\begin{equation}
\mathcal C_p(t_1;\,t_2)\,=\,\exp\left[-\frac{\pi}{4}\frac{(t_1-t_2)^2}{t_{\rm s}^2}\right]\,,
\label{eq:Cmu}
\end{equation}
in terms of the lab-frame scattering time $t_{\rm s}$, which characterizes the random walk of the particle in the configuration space. In particular, 
$\left(\mathbf{x_1}-\mathbf{x_2}\right)^2\,\simeq\,2t_{\rm s}\left\vert{t_1}-{t_2}\right\vert/3$, in Eq.~(\ref{eq:corr}).

The computation of the rhs of Eq.~(\ref{eq:diff0}) is somewhat laborious in the present case, yet essential in order to identify the respective contributions of expansion, shear, vorticity and acceleration, as well as to extract the leading factor of $\gamma_u$ in the relativistic limit. When evaluating the integrals over the correlation functions in what follows, the various coherence wave numbers $K_I$ and frequencies $\Omega_I$ are set equal to common (respective) values $K$ and $\Omega$, for ease of notation. 

\subsection{Subrelativistic limit}
Consider first the nonrelativistic limit, to lowest order in $\beta_u$, with $\gamma_u\,\approx\,1$. Then $\hat p^{\hat \i} \,\simeq\, p^i$ etc.  Computing the diffusion coefficient at fixed energy, $p(t)\,=\,p$, hence $p_1\,=\,p_2\,=\,p$ and using ${\rm d}\tau\,=\,{\rm d}t\, m/p$ to rewrite the integral in terms of (lab-frame) time leads to:
\begin{eqnarray}
\left\langle\Delta p^2\right\rangle&\,=\,&p^2\int{\rm d}t_1{\rm d}t_2\,\biggl\{\left\langle \theta^2\right\rangle \mathcal C_\theta(t_1,t_2)\left[1+\frac{2}{3}\mathcal C_p(t_1,t_2)^2\right] \nonumber\\
&&\quad\quad\quad\quad\,+\, \left\langle a^2\right\rangle \mathcal C_a(t_1,t_2)\mathcal C_p(t_1,t_2)\nonumber\\
&&\quad\quad\quad\quad\,+\,\frac{2}{5} \left\langle\sigma^2\right\rangle \mathcal C_\sigma(t_1,t_2)\mathcal C_p(t_1,t_2)^2\biggr\}\,.
\label{eq:diffs1}
\end{eqnarray}

The factor of unity that enters as the first term in the brackets with $\mathcal C_\theta$ in the prefactor deserves particular scrutiny. This term does not vanish even if the particle momentum decorrelates, meaning that it is not associated with the transport of the particle in the turbulence. It is rather associated with the average over all possible realizations of the turbulent field. According to the ergodic hypothesis, this amounts to considering a large number $N\,\gg\,1$ of particles spread over scales significantly larger than the coherence scale of the turbulent field, and taking the averaged contribution over the whole population. As indicated by Eq.~(\ref{eq:avf}), the average force is zero, but its rms is not: some particles will experience compression in some regions, and some others decompression, which broadens the distribution function and thus introduces the above term. If one is rather interested in following a single particle through a given turbulent setting, then this term should be discarded. We therefore separate these two contributions and write:
\begin{eqnarray}
\left\langle\Delta p^2\right\rangle_{N}&\,=\,&p^2\int{\rm d}t_1{\rm d}t_2\,\left\langle \theta^2\right\rangle \mathcal C_\theta(t_1,t_2)\,,\nonumber\\
\left\langle\Delta p^2\right\rangle_{1}&\,=\,&\left\langle\Delta p^2\right\rangle\,-\,
\left\langle\Delta p^2\right\rangle_{N}\,.
\label{eq:diffs1b}
\end{eqnarray}

These integrals are evaluated assuming $\Omega\,\ll\,K$, which is appropriate for a subrelativistic turbulent flow. Compact expressions are obtained in the following limits of interest. 

\subsubsection{$\Omega/K^2\,\ll\,t_{\rm s}\,\ll\,K^{-1}$}
In this limit, the particle is effectively trapped in the turbulence and it propagates diffusively over a scale of correlation $\sim\,K^{-1}$. Explicit integration of Eq.~(\ref{eq:diffs1}) then leads to:
\begin{eqnarray}
\frac{\left\langle\Delta p^2\right\rangle_N}{2\Delta t}&\,=\,&\frac{6}{\pi}\,p^2\,\frac{\left\langle\theta^2\right\rangle}{K^2  t_{\rm s}}\,,\nonumber\\
\frac{\left\langle\Delta p^2\right\rangle_1}{2\Delta t}&\,=\,&\frac{\sqrt{2}}{3}p^2\,t_{\rm s}\, \left(\left\langle \theta^2\right\rangle\,+\,
\frac{3}{5}\left\langle\sigma^2\right\rangle\,+\,
\frac{3}{\sqrt{2}}\left\langle a^2\right\rangle\right)\,,\nonumber\\
&&
\label{eq:diffs2a}
\end{eqnarray}
to lowest order in $\Omega/K^2$ and $t_{\rm s}$. Concerning the population average, indexed with $_N$, only compressive modes  contribute to order $\mathcal O(1/t_{\rm s})$; incompressible modes contribute to order $\mathcal O(t_{\rm s})$, in agreement with Ref.~\cite{1983ICRC....9..313B}.

Note that the contribution of the acceleration is well subdominant compared to other terms in the subrelativistic limit, since $\left\langle a^2\right\rangle\,\sim\,\Omega^2\,\beta_u^2$ and $\Omega\,\ll\,K$, while $\left\langle \theta^2\right\rangle\,\sim\,K^2\,\beta_u^2$.

The scattering time $t_{\rm s}$ cannot be set to arbitrarily low values because turbulent advection implies a minimum value for the spatial diffusion coefficient of the order of $\beta_u/K$, which describes transport of the particle at average velocity $\beta_u$ with mean free path $K^{-1}$. For a particle traveling at $c$, this corresponds to an effective $t_{\rm s,\,min}\,\approx\,\,\beta_u/K$. Then, for $\left\langle\theta^2\right\rangle\,\sim\,K^2\beta_u^2$,
\begin{eqnarray}
\frac{\left\langle\Delta p^2\right\rangle_N}{2\Delta t}&\,\lesssim\,&p^2\,K \beta_u\,,\nonumber\\
\frac{\left\langle\Delta p^2\right\rangle_1}{2\Delta t}&\,\gtrsim\,&p^2\,K \beta_u^3
\,.\label{eq:diffs2b}
\end{eqnarray}

\subsubsection{$\Omega/K^2\,\ll\,K^{-1}\,\ll\,t_{\rm s}$} 
Particles can now cross many coherence lengths of the turbulence before losing the memory of their initial trajectory. One derives
\begin{eqnarray}
\frac{\left\langle\Delta p^2\right\rangle_N}{2\Delta t}&\,=\,&\frac{6}{\pi}\,p^2\,\frac{\left\langle\theta^2\right\rangle}{K^2 t_{\rm s}}\,,\nonumber\\
\frac{\left\langle\Delta p^2\right\rangle_1}{2\Delta t}&\,=\,&\frac{4}{\pi}\,p^2\,\frac{1}{K^2 t_{\rm s}}
\left(\left\langle\theta^2\right\rangle\,+\,
\frac{3}{5}\left\langle\sigma^2\right\rangle\,+\,
\frac{3}{2}\left\langle a^2\right\rangle\right)\,,\nonumber\\
&&
\label{eq:diffs3}
\end{eqnarray}
in agreement with the scaling obtained in Ref.~\cite{1988SvAL...14..255P}.

\subsubsection{$t_{\rm s}\,\ll\,\Omega/K^2\,\ll\,K^{-1}$}
As in the first case, the particles are effectively trapped in a correlation cell of the turbulence. However, due to the decorrelation over time of this turbulence,  particles may nevertheless experience uncorrelated velocity fields:
\begin{eqnarray}
\frac{\left\langle\Delta p^2\right\rangle_N}{2\Delta t}&\,=\,&p^2\,\frac{\left\langle\theta^2\right\rangle}{\Omega}\,,\nonumber\\
\frac{\left\langle\Delta p^2\right\rangle_1}{2\Delta t}&\,=\,&\frac{\sqrt{2}}{3}\,p^2\,t_{\rm s}\,\left(\left\langle\theta^2\right\rangle\,+\,
\frac{3}{5}\left\langle\sigma^2\right\rangle\,+\,
\frac{3}{\sqrt{2}}\left\langle a^2\right\rangle\right)\,.\nonumber\\
&&
\label{eq:diffs4}
\end{eqnarray}
The meaning of the first term (population average) differs from that obtained before. It expresses the second moment of the momentum distribution of a large number of particles that are trapped at a fixed location in the turbulence, but averaged over a range of times that is significantly larger than the coherence time of the turbulence. Noncompressive modes contribute to order $\mathcal O(t_{\rm s})$.  Regarding the second term, it coincides with the minimum value given in Eq.~(\ref{eq:diffs2b}) if $\Omega\,\sim\,\beta_u\,K c$, as one should expect.

\subsection{Relativistic turbulence}\label{sec:relturb}
Consider now the relativistic limit $\gamma_u\beta_u\,\gg\,1$. The general expression for $\left\langle\Delta p^2\right\rangle$ is significantly longer than that for the subrelativistic limit, which was written to first order in $\beta_u^2$, and in fact too lengthy to be explicitly reproduced here. The integrand of the rhs of Eq.~(\ref{eq:diff0}) scales in proportion to $\gamma_u^2\, {\hat p_1}^2\, {\hat p_2}^2$ times products of correlation functions of compressible, shear and vortical motions, accelerations as well as $\mathcal C_p$ and $\mathcal C_u$; the subscripts $1$ and $2$ refer here and thereafter to the time at which they are computed.

In the present case, it becomes important to distinguish the diffusive transport in momentum space from the energy gain associated to the change of frame and momenta between initial and final states, which increases $p^t$ by a factor of the order of $\gamma_u^2$. To this effect, the following calculation accounts for the evolution of momenta during time in the course of integration. In the present nonresonant scheme, the diffusion coefficient is expected to scale as $p^2$ -- as in the subrelativistic limit -- because the particle is insensitive to the inner structure of the turbulence spectrum. Leaving aside this issue of energy gain at initial and final times for now, one expects a diffusion coefficient of the form $D_{pp}\,=\,p^2/t_{pp}$ with $t_{pp}$ the acceleration timescale. In stochastic resonant particle-wave interactions, $t_{pp}\,\sim\,t_{\rm s}/v_\phi^2$, where $v_\phi$ represents the typical phase velocity of the waves; in the absence of resonance, or for strongly broadened resonances, $t_{\rm s}$ becomes independent of $p$, {\it e.g.}~\cite{2012ApJ...754..103B}, so that $D_{pp}\,\propto\,p^2$. In turn, this implies $p\,\propto\,\exp\left(\Delta t/t_{pp}\right)$. Including now the first-order Fermi energy gain, one anticipates the scaling:
\begin{equation}
\left\langle \Delta p^2\right\rangle\,\sim\,\left\langle p\right\rangle^2\,\sim\,\gamma_u^4\,p_0^2\,e^{2\Delta t/t_{pp}}\,,
\label{eq:diffu1}
\end{equation}
with $p_0\,=\,p^t(0)$. This scaling will be confirmed {\it a posteriori}.

The notion of $t_{pp}$ as an acceleration timescale needs to be clarified because of the competition between the two energy gain mechanisms, first-order {\it vs} diffusive. Strictly speaking, it characterizes the rate of acceleration of the diffusive process and, as such, holds asymptotically in time. On timescales $\Delta t\,\gtrsim\,t_{\rm d}$, with $t_{\rm d}\,=\,{\rm min}\left(t_{\rm s},\,t_{\rm c}\right)$ the decorrelation time, written in terms of the scattering timescale $t_{\rm s}$ and the coherence time of the velocity field $t_{\rm c}$, the particle has already acquired a fixed $\sim\,\gamma_u^2$ energy gain even if diffusion has not taken place. Hence, on shorter timescales, the acceleration timescale should rather be understood as $t_{\rm d}/\gamma_u^2$.

The following calculation determines $t_{pp}$ by computing $\left\langle \Delta p^2\right\rangle$ as a function of time, and by expliciting the relationship $\left\langle \Delta p^2\right\rangle\,\simeq\,\left\langle p\right\rangle^2$. To carry out the time integrals in Eq.~(\ref{eq:diff0}), the following scaling is used: ${p}_{1,2}\,=\,\gamma_u^2\, p(0)\, \exp\left(t_{1,2}/t_{pp}\right)$ corresponding to $\hat p^{\hat 0}_{1,2}\,=\,\gamma_u\, p(0)\, \exp\left(t_{1,2}/t_{pp}\right)$ with $t_{1,2}$, $t_{\rm s}$ and $t_{pp}$ defined in $\mathcal R_{\rm L}$. The correlation function for the momenta is defined as in the subrelativistic limit, and similarly for the velocity field correlation function. The integrals can be carried out using the changes of variables ${\rm d}\tau_{1,2}\,=\,{\rm d}t_{1,2}/(p_{1,2}/m)$, then $\sigma_t\,=\,(t_1+t_2)/2$ and $\Delta_t\,=\,(t_1-t_2)$ as usual, integrating $\sigma_t$ from $0$ to $\Delta t$, which as before is assumed to be much larger than $t_{\rm s}$. The remaining integral over $\Delta_t$ converges to a constant value as $\Delta_t\,\rightarrow\,+\infty$, as usual for diffusion processes. Eventually, one obtains a lengthy expression which takes the form
\begin{equation}
\left\langle \Delta p^2\right\rangle\,\simeq\,\left\langle p\right\rangle^2\, t_{pp}\,A_u\left(t_{\rm s},K,\,\Omega\right)\,,
\label{eq:accu}
\end{equation}
with $A_u\left(t_{\rm s},K,\,\Omega\right)$ to be specified further on, and $\left\langle p\right\rangle^2\,=\,\gamma_u^4\,p_0^2\,\exp\left(2\Delta t/t_{pp}\right)$, which allows to identify the diffusive acceleration timescale as
\begin{equation}
t_{pp}\,\sim\,A_u^{-1}\,.
\label{eq:accub}
\end{equation}

As in the subrelativistic limit, one needs to distinguish the contribution from the terms that do not vanish in the limit $C_p(t_1;\,t_2)\rightarrow 0$, and that correspond to averages taken over a population of particles experiencing all possible realizations of the turbulence, from terms that vanish in the limit $C_p(t_1;\,t_2)\rightarrow 0$, and that characterize the effect of an average turbulence experienced by one particle. The former gives a contribution written $t_{pp}^{(N)}$ and the latter, $t_{pp}^{(1)}$, following earlier notations.

Making explicit the various $A_u(t_{\rm s},\,K,\Omega)$ functions, one obtains the following approximations, to lowest order in the $t_{\rm s}K$ expansion,
\medskip

i) In the small scattering timescale $Kt_{\rm s}\,\ll\,1$ and stationary $\Omega\,=\,0$ limit:
\begin{eqnarray}
t_{pp}^{(N)}&\,\simeq\,&\frac{K^2 t_{\rm s}}{1.23\left\langle\theta^2\right\rangle
\,+\,0.14\,\left\langle\sigma^2\right\rangle
\,+\,0.73\,\left\langle a^2\right\rangle
}\,,\nonumber\\
t_{pp}^{(1)}&\,\simeq\,& \frac{t_{\rm s}^{-1}}{0.87\left\langle\theta^2\right\rangle
\,+\,0.40\,\left\langle\sigma^2\right\rangle
\,+\,0.52\,\left\langle\omega^2\right\rangle
\,+\,2.4\,\left\langle a^2\right\rangle
}
\,.\nonumber\\
&&
\label{eq:diffu2a}
\end{eqnarray}
\medskip

ii) In the small scattering timescale limit $Kt_{\rm s}\,\ll\,1$ but $\Omega\,=\,K$:
\begin{eqnarray}
t_{pp}^{(N)}&\,\simeq\,&\frac{K}{0.38\left\langle\theta^2\right\rangle
\,+\,0.07\,\left\langle\sigma^2\right\rangle
\,+\,0.28\,\left\langle a^2\right\rangle
}\,,\nonumber\\
t_{pp}^{(1)}&\,\simeq\,& \frac{t_{\rm s}^{-1}}{0.87\left\langle\theta^2\right\rangle
\,+\,0.40\,\left\langle\sigma^2\right\rangle
\,+\,0.52\,\left\langle\omega^2\right\rangle
\,+\,2.4\,\left\langle a^2\right\rangle
}
\,.\nonumber\\
&&
\label{eq:diffu2b}
\end{eqnarray}
\medskip

iii) In the large scattering timescale limit $K t_{\rm s}\,\gg\,1$, independently of $\Omega/K$:
\begin{eqnarray}
t_{pp}^{(N)}&\,\simeq\,& \frac{K^2 t_{\rm s}}{1.23\left\langle\theta^2\right\rangle
\,+\,0.14\,\left\langle\sigma^2\right\rangle
\,+\,0.73\,\left\langle a^2\right\rangle
}\,,\nonumber\\
t_{pp}^{(1)}&\,\simeq\,& \frac{K^2t_{\rm s}}{1.9\left\langle\theta^2\right\rangle
\,+\,0.94\,\left\langle\sigma^2\right\rangle
\,+\,0.90\,\left\langle\omega^2\right\rangle
\,+\,7.3\,\left\langle a^2\right\rangle
}
\,.\nonumber\\
&&
\label{eq:diffu2c}
\end{eqnarray}

Interestingly, the vorticity contributes in about equal amounts to shear, expansion and acceleration in this ultrarelativistic limit. If $\langle\theta^2\rangle\,\sim\,K^2\langle u^2\rangle$ and similarly for the other components, then the diffusive acceleration timescale can be written more simply as
\begin{equation}
t_{pp}\,\sim\,\langle u^2\rangle^{-1}\begin{cases}\left(K^2\,t_{\rm s}\right)^{-1}\, & \quad \left(K t_{\rm s}\,\ll\,1\right)\\
t_{\rm s}\, & \quad\left(K t_{\rm s}\,\gg\,1\right)\,.\end{cases}
\label{eq:tppur}
\end{equation}
Noting that $\langle u^2\rangle\,\simeq\,\gamma_u^2$, this offers the possibility of fast acceleration if the turbulence is truly relativistic.

\subsection{Discussion}
The present calculations cover a larger domain of validity than other methods. They do not, in particular, rely on a perturbative expansion in powers of the electric field, as the quasilinear calculation, which is thus limited to the lowest order in $\beta_u^2$. The present calculations provide results to order $u^2$ (where $u\,=\,\beta_u\gamma_u$), which actually derive from higher-order terms through the reduction of these latter in two-point functions.  The diffusion coefficient here obtained cannot either be extracted from a fundamental cosmic-ray transport equation, because of the nonuniformity of the flow.

The present calculations, in both the sub- and ultrarelativistic limits, assume a monochromatic turbulence with a well-defined coherence frequency $\Omega$ and wave number $K$. They could however be generalized to the case of a broad turbulence spectrum, extending over a range of wave numbers and frequencies encompassing the scale $t_{\rm s}^{-1}$, using the scalings given for $K t_{\rm s}\,\gg\,1$ and $K t_{\rm s}\,\ll\,1$. 

These scalings can be understood as follows: in the limit $K t_{\rm s}\,\ll\,1$, the particle travels in a fully diffusive manner on all scales of interest; hence, it takes a time $K^{-2}/t_{\rm s}$ for the particle to cross a coherence cell of the turbulence, and on such timescales, it gains a factor $\langle u^2\rangle$ in energy. In contrast, if $K t_{\rm s}\,\gg\,1$, the particle now travels mostly in a ballistic manner over the coherence length scale of the turbulence; hence, it now takes a timescale $t_{\rm s}$ for the particle to suffer a deflection of the order of unity to obtain the same energy gain. If the scattering of particles is governed by nonresonant processes, {\it e.g.} random field line diffusion, one should expect $t_{\rm s}\,\sim\, K^{-1}$, in which case both limits would lead to the same scaling $t_{pp}\,\sim\,\langle u^2\rangle^{-1} K^{-1}$.

\section{Flows with nonzero mean velocity}\label{sec:nonz}
Assuming that the scattering timescale is much smaller than the scale of variation of the flow allows to develop perturbatively the distribution function of the accelerated particles and to derive an approximate transport equation~\cite{1967PhFl...10.2620H,1968Icar....8...54D,1972ApJ...172..319J,1975MNRAS.172..557S, 1985ApJ...296..319W,1989ApJ...336..243S,1993PhyU...36.1020B, 1993ApJ...405L..79W,2018MNRAS.479.1747A}. Such transport equations then provide the diffusion coefficient in terms of the four-acceleration $a^\alpha\,=\,u^\beta\,{u^\alpha}_{;\beta}$ and the shear tensor ${\sigma^{\scriptscriptstyle{ \rm 4D}}}_{\alpha\beta}$. 

This section illustrates the use of the present approach to calculate such diffusion coefficients and compares them with existing results in known cases. One advantage of the present method is that it is not limited to the approximation $t_{\rm s}\,\ll\,\vert u/\partial u\vert$.

\subsection{Shear acceleration}
As a first example, consider the physics of shear acceleration, in which a particle gains energy provided it can interact at various points of a sheared velocity field, {\it e.g.}~\cite{2006ApJ...652.1044R,1981SvAL....7..352B,*1988ApJ...331L..91E,*1990A&A...238..435O,*1990ApJ...356..255J, *2002A&A...396..833R,2004ApJ...617..155R,*2005ApJ...632L..21R,*2016ApJ...833...34R, *2017ApJ...842...39L,*2018ApJ...855...31W}.
In these references, the acceleration term has been derived either from a transport equation that accounts for the relevant inertial corrections or from a microscopic analysis, where one follows the energy gain through local Lorentz transforms from one rest frame to another. The present approach generalizes the latter and avoids dealing with the complexities of the fully relativistic transport equation.

\subsubsection{Shear acceleration in a plane}\label{sec:shp}
Consider for instance the simplest case of a two-dimensional shear flow in Cartesian coordinates and in flat space-time, where the only nonzero spatial component of the velocity is $u^x(y)$ and the only nonzero derivative ${u^x}_{,y}$. Strictly speaking, this flow does not describe a pure shear, as it contains an equal amount of vorticity. 

In the relativistic limit, one must carefully define the lab frame. Here, an obvious choice is to boost this lab frame to the comoving frame at one point $y_0$ of the flow, so that the above shear profile is redefined in such a way that $u^x(y_0)\,=\,0$. It proves convenient to define $y_0$ as that where the trajectory starts and to assume that the trajectory comes back to this point at the final time. Then one can dispense with the extra boost at initial and final times in Eq.~(\ref{eq:diff1}). 

We thus seek to evaluate:
\begin{equation}
  D_{pp}\,\underset{\Delta t\,\rightarrow\,+\infty}=\,\frac{1}{2\Delta t}\int{\rm d}\tau_1{\rm d}\tau_2\,\left\langle\frac{{\rm d}{\hat p}^{\hat 0}}{{\rm d}\tau_1}\frac{{\rm d}{\hat p}^{\hat 0}}{{\rm d}\tau_2}\right\rangle \,.
\label{eq:she1}
\end{equation}
The nonzero connection terms are: $\widehat \Gamma^{\hat 0}_{\hat 1\hat 2}\,=\,\widehat \Gamma^{\hat 1}_{\hat 0\hat 2}\,=\,{u^x}_{,y}/\gamma_u$. Consequently,
\begin{eqnarray}
\left\langle\frac{{\rm d}{\hat p}^{\hat 0}}{{\rm d}\tau_1}\frac{{\rm d}{\hat p}^{\hat 0}}{{\rm d}\tau_2}\right\rangle &\,=\,& \frac{{u^x}_{,y}(\tau_1){u^x}_{,y}(\tau_2)}{\gamma_{u}(\tau_1)\gamma_{u}(\tau_2)}\,,\nonumber\\
&&\quad\times \left\langle \hat p^{\hat 1}(\tau_1)\, \hat p^{\hat 2}(\tau_1)\, \hat p^{\hat 1}(\tau_2)\,\hat p^{\hat 2}(\tau_2)\right\rangle\,.\nonumber\\
&&
\label{eq:she2}
\end{eqnarray}
The above calculation does not include any stochastic term in the velocity field, even though turbulence must be present to guarantee the transport of particles across the shear. In so doing, it follows usual shear acceleration studies. It is possible however, to use the calculations of the previous section and add in these nonzero turbulent fluctuations. We further assume that the scattering timescale $t_{\rm s}$ does not depend on $y$, even though it might if the magnetic field itself exhibits a spatially dependent profile.

On timescales $\Delta t\,\gg\,t_{\rm s}$, where $t_{\rm s}$ is here as well specified in the lab frame, $\hat p^{\hat 1}$ and $\hat p^{\hat 2}$ are uncorrelated; hence, $\left\langle \hat p^{\hat 1}(\tau_1)\, \hat p^{\hat 2}(\tau_1)\, \hat p^{\hat 1}(\tau_2)\,\hat p^{\hat 2}(\tau_2)\right\rangle\,=\,C_p(t_1;\,t_2)^2\hat p^4/9$.

In the limit $t_{\rm s}\,\ll\,L_{\rm shear}$ with $L_{\rm shear}\,=\,\left\vert {u^x}_{,y}(y_0)\right\vert^{-1}$ the shear length scale, one can safely assume that the particle only explores the immediate vicinity of the point $y_0$, so that the spatial evolution of the particle can be neglected. Using ${\rm d}\tau_{1,2}\,=\,{\rm d}t_{1,2}\,m/p_{1,2}$ as before, the integral in Eq.~(\ref{eq:she2}) is straightforward, and one derives
\begin{equation}
D_{pp}\,=\,\frac{1}{9\sqrt{2}}\,p^2\,\left[{u^x}_{,y}(y_0)\right]^2 t_{\rm s}\,,
\label{eq:she3}
\end{equation}
which coincides with the standard result up to a factor of the order of unity~\cite{2004ApJ...617..155R}.

In the limit $t_{\rm s}\,\gg\,L_{\rm shear}$, the particle can now explore the shear profile. Since the shear cannot extend indefinitely, it is reasonable to assume that $L_{\rm shear}$ also characterizes the extent of this shear profile in the $y-$direction, meaning that $u^x$ reaches a constant on larger distance scales. It  then suffices to describe the shear profile on large length scales as a discontinuity at $y\,=\,y_0$, {\it viz.} ${u^x}_{,y}\,=\,\Delta {u^x}\,\delta(y - y_0)$. To model the trajectory of the particle, we assume that $y_1$ at $t_1\,<\,t_2$ can take any value around $y_0$ within $\pm\sqrt{t_{\rm s}t_1}$, and that the probability of finding the particle at $y_2$ at $t_2$ is tied to the position at $t_1$ through the standard diffusive propagator in 1D, with spatial diffusion coefficient $D_{yy}\,\simeq\,t_{\rm s}/3$. The average in Eq.~(\ref{eq:she2}) then becomes
\begin{equation}
\left\langle\frac{{\rm d}{\hat p}^{\hat 0}}{{\rm d}\tau_1}\frac{{\rm d}{\hat p}^{\hat 0}}{{\rm d}\tau_2}\right\rangle \,\propto\, \frac{\hat p^4}{m^2}\,\left(\Delta {u^x}\right)^2\frac{\exp\left[-\frac{(t_1-t_2)^2}{t_{\rm s}^2}\right]}
{\sqrt{t_1 t_{\rm s}}\sqrt{D_{yy}(t_2-t_1)}}\,.
\label{eq:she4}
\end{equation}
Interestingly, the resulting integral over ${\rm d}t_1{\rm d}t_2$ does not reveal a diffusive behavior: if it were so, this integral should scale as $\Delta t$ in the limit $\Delta t\,\rightarrow\,+\infty$ ($\Delta t$ corresponds to the upper bound of integration on $t_1$ and $t_2$) while the above integral rather scales as $1/\sqrt{\Delta t}$. The acceleration is no longer diffusive here, because particles can escape the shear profile on either side $y\,>\,0$ or $y\,<\,0$.  Hence, shear acceleration stops once the particle reaches an energy such that $t_{\rm s}\,\sim\,L_{\rm shear}$.

For the sake of illustration, consider nevertheless the case in which particles remain confined in a box of size $t_{\rm s}$ in the $y-$ direction, which forces the particles to experience the shear. There is no net mean energy gain here, contrary to the case of shock acceleration, because this mean energy gain scales as $\langle \hat p^{\hat 1}(t)\hat p^{\hat 2}(t)\rangle$, which vanishes in the diffusion approximation. Diffusion nevertheless occurs, and repeating the above calculation, now assuming that $y_1$ and $y_2$ can take any value in the interval $\pm t_{\rm s}$, one obtains
\begin{equation}
D_{pp}\,\sim\,\frac{1}{18\sqrt{2}}\,p^2\,\frac{\left(\Delta u^x\right)^2}{t_{\rm s}}\,,
\label{eq:she5}
\end{equation}
so that the acceleration timescale $t_{pp}\,\propto\,t_{\rm s}/\left(\Delta u^x\right)^2$.

\subsubsection{Compressive or expansive flow}
As a further example of interest, consider a flow undergoing deceleration and/or acceleration in one dimension, with velocity profile: $u^\mu\,=\,\left[\gamma_u(x),u^x(x),0,0\right]$, in Cartesian coordinates (and flat space-time). Such a compressive flow provides an obvious model for the precursor of a shock front, on microscopic scales that resolve the shock transition. The nonzero components of the connection are: $\widehat\Gamma^{\hat 0}_{\hat 1\hat 0}\,=\,\widehat\Gamma^{\hat 1}_{\hat 0\hat 0}\,=\,\beta_u {u^x}_{,x}$ and $\widehat\Gamma^{\hat 0}_{\hat 1\hat 1}\,=\,\widehat\Gamma^{\hat 1}_{\hat 0\hat 1}\,=\,{u^x}_{,x}$, giving
\begin{equation}
\frac{{\rm d}{\hat p}^{\hat 0}}{{\rm d}\tau}\,=\,-{u^x}_{,x}\,\frac{\hat p^{\hat 1}}{m}\left(\beta_u \hat p^{\hat 0} + \hat p^{\hat 1}\right)\,,
\label{eq:she9}
\end{equation}
which has a nonzero mean value in the diffusion approximation,
\begin{equation}
\left\langle\frac{{\rm d}{\hat p}^{\hat 0}}{{\rm d}\tau}\right\rangle\,=\,-\frac{1}{3}\frac{\hat p^2}{m}{u^x}_{,x}\,.
\label{eq:she10}
\end{equation}
One recovers here the energy gain or loss due to compression or expansion along one spatial direction, assuming isotropization of the momentum in the three directions, {\it e.g.} \cite{1985ApJ...296..319W,*1989ApJ...340.1112W}.

The diffusion term can be obtained after subtracting properly the square of this mean energy gain or loss, assuming that $t_{\rm s}\,\ll\,\vert {u^x}_{,x}/\gamma_u\vert^{-1}$
\begin{align}
\underset{\Delta t\,\rightarrow\,+\infty}{\rm lim}&\frac{1}{2\Delta t}\int{\rm d}\tau_1{\rm d}\tau_2\,\biggl\{
\left\langle\frac{{\rm d}{\hat p}^{\hat 0}}{{\rm d}\tau_1}\frac{{\rm d}{\hat p}^{\hat 0}}{{\rm d}\tau_2}\right\rangle \,-\,\left\langle\frac{{\rm d}{\hat p}^{\hat 0}}{{\rm d}\tau_1}\right\rangle\left\langle\frac{{\rm d}{\hat p}^{\hat 0}}{{\rm d}\tau_2}\right\rangle\biggr\}\nonumber\\
\,=\,&\frac{\sqrt{2}-3\beta_u^2}{9}\,\hat p^2\,\left({u^x}_{,x}\right)^2\,t_{\rm s}\,.
\label{eq:she11}
\end{align}
This equation takes the same general form as the diffusion coefficient for shear acceleration. These results describe in particular the diffusion in energy of a particle in a shock transition, with a mean free path that is shorter than the width of the shock transition, {\it i.e.}, the injection process in the shock transition layer before the particle has reached an energy such that $t_{\rm s}\,\gg\,\vert u^x_{,x}/\gamma_u\vert^{-1}$. This general scaling for both the mean energy gain and the diffusion coefficient has been derived through a Vlasov-Fokker-Planck model in Ref.~\cite{2019PhRvE.100c3209L}.

\section{Flows in nontrivial geometries}\label{sec:geom}
nontrivial geometries imply a nontrivial ${e{_{\rm L}}^{\overline a}}_\mu$ vierbein, but the techniques developed in the previous paragraphs remain unchanged. Here it proves useful to define
\begin{equation}
\Delta^{\hat 0\hat 0}\,=\,\int{\rm d}\tau_1{\rm d}\tau_2\,\biggl\{
\left\langle\frac{{\rm d}{\hat p}^{\hat 0}}{{\rm d}\tau_1}\frac{{\rm d}{\hat p}^{\hat 0}}{{\rm d}\tau_2}\right\rangle \,-\,\left\langle\frac{{\rm d}{\hat p}^{\hat 0}}{{\rm d}\tau_1}\right\rangle\left\langle\frac{{\rm d}{\hat p}^{\hat 0}}{{\rm d}\tau_2}\right\rangle\biggr\}\,,
\label{eq:Itt}
\end{equation}
which characterizes the diffusion rate (around the mean energy gain) in the locally inertial frame. 

\subsection{Centrifugal and shear acceleration in a disk}
Consider for instance the problem of shear acceleration in cylindrical coordinates $(t,r,\phi,z)$, for a circular flow profile $u^\mu\,=\,\{u^t(r),0,u^\phi(r),0\}$ with angular velocity $\Omega(r)\,=\,u^\phi(r)/u^t(r)$. The explicit $r-$dependence of $g_{\phi\phi}\,=\,r^2$ renders the lab vierbein nontrivial: ${{e_{\rm L}}^{\overline a}}_\mu\,=\,{\rm diag}\left(1,1,r,1\right)$. 

The scattering time $\hat t_{\rm s}$ is naturally defined in the locally inertial frame, since the turbulence that leads to the scattering is carried by the flow along its circular orbit. One derives
\begin{equation}
\frac{{\rm d}\hat p^{\hat 0}}{{\rm d}\tau}\,=\,-r\frac{\hat p^{\hat 1}}{m}
\frac{\hat p^{\hat 0}\Omega^2\,-\,\hat p^{\hat 2}\Omega_{,r}}{1-r^2\Omega^2}\,,
\label{eq:cylf}
\end{equation}
so that its average vanishes if $\hat p^{\hat \i}$ is randomized on the unit sphere, and
\begin{eqnarray}
\left\langle\Delta p^t\,\Delta p^t\right\rangle&\,\underset{\Delta t\,\rightarrow\,+\infty}=\,& {u^t}^2
\left[1+\frac{1}{3}r^2\Omega^2\right]\,\Delta^{\hat 0\hat 0}\,,
\label{eq:shd1}
\end{eqnarray}
with
\begin{equation}
\Delta^{\hat 0\hat 0}\,=\,2\widehat{\Delta t}\,\frac{\hat p^2 r^2 {\hat t_{\rm s}}}{18}\,
\frac{6 \Omega ^4\,+\,\sqrt{2} {\Omega_{,r}}^2}{\left(1-r^2 \Omega^2\right)^2}\,,
\label{eq:shd2}
\end{equation}
with the time interval $\widehat{\Delta t}$ defined in the locally inertial frame.
The diffusion coefficient contain both a shear contribution, that scales as ${\Omega_{,r}}^2$, and a centrifugal term $\propto\,\Omega^4$, in agreement with Ref.~\cite{2002A&A...396..833R}.

Following the argument of Sec.~\ref{sec:relturb} for relativistic motion and a scattering time $\hat t_{\rm s}$ independent of $\hat p^{\hat 0}$, the momentum in the lab frame evolves as: $\left\langle p\right\rangle^2\,\sim\,\left\langle \Delta p^2\right\rangle\,\sim\,{u^t}^4\,p_0^2\,\exp\left(2\widehat{\Delta t}/\hat t_{pp}\right)$ with $\hat t_{pp}\,=\,2{\widehat{\Delta t}}\,{{\hat p^{\hat 0}}\,}^2/
\Delta^{\hat 0\hat 0}$ and $\widehat{\Delta t}\,=\,\Delta t/u^t$.

\subsection{Stochastic unipolar induction}
A rotating conductor carrying a magnetic field sets up a potential difference between the pole (where $\mathbf{E}\,=\,0$) and the equator (where $\mathbf{E} \,=\,-\boldsymbol{\beta}\times\mathbf{B}$). This gives rise to unipolar acceleration if the particles are able to experience the voltage drop, or a fraction of it. 

In these scenarios, how the particles travel across the magnetic field lines is usually left unspecified. In the presence of scattering and turbulence, however, cross-field transport is guaranteed. In a generic configuration, the magnetic field winds up so that at large radii from the central object, it is mostly toroidal while the motion is mostly radial. Thus, assuming $u^\mu\,=\,\left\{u^t(r,\theta),u^r(r,\theta),0,0\right\}$ in spherical coordinates $(t,r,\theta,\phi)$, one obtains
\begin{equation}
\frac{{\rm d}\hat p^{\hat 0}}{{\rm d}\tau}\,=\,
-\frac{1}{3}\frac{\hat p^2}{m} \boldsymbol{\nabla}\cdot\mathbf{u}\,,
\label{eq:sph1}
\end{equation}
as expected. Its average no longer vanishes, guaranteeing that particles cool as the wind accelerates and conversely. Furthermore,
\begin{align}
\Delta^{\hat 0\hat 0}\,=\,&2\widehat{\Delta t}\,\frac{\hat p^2 {\hat t_{\rm s}}}{18r^2}\,
\biggl[4{u^r}^2+2\left(r {u^r}_{,r}\right)^2
\,+\, \frac{{{u^r}_{,\theta}}^2}{{u^t}^2}
\,+\, 6 \frac{\left(r {u^r}_{,r}\right)^2}{{u^t}^2}\biggr]\,.
\label{eq:sph2}
\end{align}
The acceleration now comprises a shear along the direction of motion as well as a shear contribution transverse to this direction, through the $\theta-$dependence.
As for the previous case of centrifugo-shear acceleration, the quantity $\Delta^{\hat 0\hat 0}/\left[2\widehat{\Delta t}\right]$ defines the diffusion coefficient $D_{\hat p\hat p}$ in the locally inertial frame, from which one can derive the acceleration timescale $\hat t_{pp}\,=\,\hat p^2/D_{\hat p\hat p}$, and, hence, the general scaling of $\langle p\rangle^2$ in the lab frame.

It may be useful to stress that the above calculations assume 
$t_{\rm s}\,\ll\,\left\vert u/\partial u\right\vert$, {\it i.e.}, that the particles scatter on a timescale much shorter than the evolution time scale of the velocity flow. 

\subsection{Fermi-type acceleration close to a black hole}
As a final example, consider the same problem of centrifugo-shear acceleration in an acceleration disk, now taking into account the presence of a central black hole. We assume here a nonrotating black hole described by the Schwarzschild metric in spherical coordinates $(t,r,\theta,\phi)$ and leave to future work the case of a rotating (Kerr) black hole:\linebreak
$g_{\mu\nu}\,=\,{\rm diag}\,\left\{-\left[1-r_{\rm H}/r\right],\left[1-r_{\rm H}/r\right]^{-1},r^2,r^2\sin^2\theta\right\}$, 
with $r_{\rm H}\,=\,2 GM/c^2$ the horizon radius in terms of the black hole mass $M$. Assume a circular orbit $u^\mu\,=\,\{u^t(r,\theta),0,0,u^\phi(r,\theta)\}$. The angular momentum of the flow is written $\ell_u\,=\,u_\phi/u_t\,=\,r^2\sin^2\theta\,\Omega/(1-r_{\rm H}/r)$ in terms of the angular velocity $\Omega\,=\,u^\phi/u^t$.

On the equator ($\theta\,=\,\pi/2$), the nonzero time components of the connection are
\begin{align}
\widehat\Gamma^{\hat 0}_{\hat 1\hat 0}&\,=\,\frac{r^3 r_{\rm H}-2 \ell_u^2 (r-r_{\rm H})^2}{2 r^{3/2} \sqrt{r-r_{\rm H}} \left[r^3-\ell_u^2 (r-r_{\rm H})\right]}\,,\nonumber\\
\widehat\Gamma^{\hat 0}_{\hat 1\hat 3}&\,=\,-\frac{\ell_u\left(2r-3r_{\rm H}\right)}
{2\left[r^3-\ell_u^2(r-r_{\rm H})\right]}\,,\nonumber\\
\widehat\Gamma^{\hat 0}_{\hat 3\hat 1}&\,=\,-\frac{\ell_u\left(2r-3r_{\rm H}\right)}
{2\left[r^3-\ell_u^2(r-r_{\rm H})\right]}\,,
\label{eq:connS}
\end{align}
so that
\begin{align}
\frac{{\rm d}\hat p^{\hat 0}}{{\rm d}\tau}&\,=\,
- \frac{\hat p^{\hat 1}}{2m\left[r^3-\ell_u^2 (r-r_{\rm H})\right]}\nonumber\\
&\quad\times\left\{2  {\hat p}^{\hat 3} \ell_u(2 r-3 r_{\rm H})-{\hat p}^{\hat 0}\frac{\left[r^3 r_{\rm H}-2 \ell_u^2 (r-r_{\rm H})^2\right]}{r^{3/2} \sqrt{r-r_{\rm H}} }\right\}\,,
\label{eq:S1}
\end{align}
whose average over the directions of $\hat p^{\hat i}$ vanishes. One thus obtains the diffusion term
\begin{align}
\Delta^{\hat 0\hat 0}&\,=\,2\widehat{\Delta t}\,
\frac{\hat p^2 \hat t_{\rm s}}{36r^3 (r-r_{\rm H}) \left[\ell_u^2 (r_{\rm H}-r)+r^3\right]^2}\nonumber\\
&\quad\times\biggl\{
12 \ell_u^4 (r-r_{\rm H})^4+3 r^6 r_{\rm H}^2+ 2\ell_u^2 r^3 (r-r_{\rm H}) \nonumber\\
&\quad\quad\times\left[4 \sqrt{2} r^2-6 r \left(2 \sqrt{2} r_{\rm H}+r_{\rm H}\right)+\left(9 \sqrt{2}+6\right) r_{\rm H}^2\right]\biggr\}\,.
\label{eq:S2}
\end{align}
In the limit $r_{\rm H}\,\rightarrow\,0$, two terms remain, one proportional to $\ell_u^4$ (the centrifugal part), and the other proportional to $\ell_u^2$ (the shear part). The present calculation assumes a constant angular momentum; hence, to recover in that flat limit the centrifugo-shear diffusion coefficient of Eq.~(\ref{eq:shd2}), one must use the substitutions $\ell_u^4\,=\,r^4\Omega^2$ and $\ell_u^2\,\rightarrow\,r^6{\Omega_{,r}}^2/4$.

The gravity field of the black hole imposes a new term, proportional to $r_{\rm H}^2$ in the bracket, which characterizes the diffusion rate of the particle energy as it scatters inwards and outwards of the potential well.

\section{Summary -- conclusions}\label{sec:conc}
The original Fermi process~\cite{1949PhRv...75.1169F} of charged particle acceleration in highly conducting plasmas has been declined into a number of variants, which all rely on the same guiding principle: energy gain takes place through the repeated interaction of the particle with inhomogeneous or time-dependent magnetic structures. The space-time dependence of the magnetic field configuration is to guarantee the absence of a global frame in which the electric field is everywhere vanishing, of course.

Building on this latter observation, the present work has described Fermi acceleration as a continuous journey of the momentum of the particle through the instantaneous frames of rest of the plasma, in which the electric field is locally screened out. In a general relativistic formulation, these local frames correspond to the {\emph comoving locally inertial frames} that can be set up at each point of the particle trajectory in configuration space. At each point, in each locally inertial frame, the particle energy evolves solely through the inertial correction that results from the space-time dependence of this inertial frame.  One advantage of the present formulation is to trade the space-time dependence of the electromagnetic field for the space-time dependence of the flow, which in various cases turns out to be better characterized. In this formulation, therefore, the Lorentz force only contributes to pitch angle scattering, which generically leads to a random walk in configuration space. Once the flow properties are defined, combining this random walk in configuration space with the law of evolution of the particle momentum in the local inertial frame allows one to determine the main characteristics of the acceleration process, in particular the mean 4-momentum change $\left\langle \Delta p^\alpha/\Delta t\right\rangle$ and its second moment $\left\langle \Delta p^\alpha\Delta p^\beta/\Delta t\right\rangle$.

One advantage of the present description lies in its general relativistic formulation, which makes its applicable to relativistic flows, to flows with nontrivial velocity patterns, or even to flows in complex geometries and/or in non-Cartesian systems. This point has been made manifest in the present paper through explicit calculations of the acceleration rate for particles subject to nonresonant turbulent acceleration, or to centrifugal-shear acceleration in an accretion disk, possibly close to the horizon of a black hole. Another advantage is the unique principle of the calculation, which allows one to describe the variants of Fermi acceleration in a similar way. Here as well, explicit examples have been presented for turbulent acceleration, shear acceleration and shock acceleration, showing that the known features of these cases can be recovered in a rather straightforward way.

An interesting lesson that emerges from these calculations is that, modulo relativistic effects tying the locally inertial and global frame between initial and final states, the acceleration timescale generically scales as $L^2/t_{\rm s}$ if $t_{\rm s}\,\ll\,L$, where $L$ characterizes the length scale of variation of the velocity, and as $t_{\rm s}$ in the opposite limit $t_{\rm s}\,\gg\,L$. The former is a notable feature of shear acceleration, while the latter occurs in shock acceleration, for instance. This result finds a simple interpretation in terms of the time it takes the particle to travel across the length scale $L$, either diffusively in space if $t_{\rm s}\,\ll\,L$, or rectilinearly, if $t_{\rm s}\,\gg\,L$.

Finally, although the present calculations have been limited to simple usual approximations, such as isotropic correlation functions of the random velocity flow or of the momenta, they can be extended to more realistic cases, at the price of increased complexity of course.
\bigskip

\begin{acknowledgments}
 This research has been supported by the ANR-14-CE33-0019 MACH project. It is a pleasure to thank A. Bykov, F. Rieger and S. Saadi for useful discussions.
\end{acknowledgments}

\bigskip

\appendix

\section{Connection terms}\label{app:conn}
In the particular case of a Cartesian coordinate basis, in flat space-time, the connection coefficients in the locally inertial frame can be written as:
\begin{align}
\widehat\Gamma^{\hat 0}_{{\hat 0}{\hat 0}}&\,=\,0\,,\nonumber\\
\widehat\Gamma^{\hat 0}_{{\hat 0}{\hat \i}}&\,=\,0\,,\nonumber\\
\widehat\Gamma^{\hat 0}_{{\hat \i}{\hat 0}}&\,=\,-\frac{u_i u^\alpha u_k  {u^k}_{,\alpha}}{\gamma_u(1+\gamma_u)} \,+\, u^\alpha u_{i,\alpha}\,,\nonumber\\
\hat\Gamma^{\hat 0}_{{\hat \i}{\hat \j}}&\,=\,-\frac{u_iu_j}{\gamma_u(1+\gamma_u)}\left(u_k {u^k}_{,t}\,+\,\frac{u^k u_l {u^l}_{,k}}{1+\gamma_u}\right)\,+\,\, u_ju_{i,t}\nonumber\\
&\quad\quad +u_{i,j}+\frac{u_ju^ku_{i,k}}{1+\gamma_u}-\frac{u_iu^ku_{k,j}}{\gamma_u(1+\gamma_u)}\,,\nonumber\\
\widehat\Gamma^{\hat \i}_{{\hat 0}{\hat 0}}&\,=\, -\frac{u_i u^\alpha u_k  {u^k}_{,\alpha}}{\gamma_u(1+\gamma_u)} \,+\, u^\alpha u_{i,\alpha}\,,\nonumber\\
\label{eq:gtab2}
\widehat\Gamma^{\hat \i}_{{\hat 0}{\hat \j}}&\,=\, {u^i}_{,j}+u_j{u^i}_{,t} -\frac{u^iu_j}{\gamma_u(1+\gamma_u)}\left(u^ku_{k,t}+ \frac{u^ku^lu_{l,k}}{1+\gamma_u}\right)\nonumber\\
&\quad\quad -\frac{u^iu^ku_{k,j}}{\gamma_u(1+\gamma_u)}+ \frac{u_ju^k{u^i}_{,k}}{1+\gamma_u}\,,\nonumber\\
\widehat\Gamma^{\hat \i}_{{\hat \j}{\hat 0}}&\,=\,-\frac{u^\alpha}{1+\gamma_u}\left(u^iu_{j,\alpha}-u_j{u^i}_{,\alpha}\right)\,,\nonumber\\
\hat\Gamma^{\hat \i}_{{\hat k}{\hat \j}}&\,=\, -\frac{1}{1+\gamma_u}\biggl[u_j\left(u^iu_{k,t}-u_k{u^i}_{,t}\right)\,+\, u^iu_{k,j}-u_k{u^i}_{,j}\nonumber\\
&\quad\quad\quad\quad\quad\quad+\frac{u_ju^lu^iu_{k,l}}{1+\gamma_u}-\frac{u_ju^lu_k{u^i}_{,l}}{1+\gamma_u}\biggr]\,.
\end{align}
As discussed in Ref.~\cite{1985ApJ...296..319W}, this connection obeys certain properties and symmetries, which notably guarantee $\widehat\Gamma^{\hat 0}_{{\hat 0}{\hat a}}\,=\,0$, $\widehat\Gamma^{\hat \i}_{{\hat \j}{\hat a}}\,=\,-\widehat\Gamma^{\hat \j}_{{\hat \i}{\hat a}}$ and $\widehat\Gamma^{\hat \i}_{{\hat 0}{\hat a}}\,=\,\widehat\Gamma^{\hat 0}_{{\hat \i}{\hat a}}$. These properties are manifest in the above equations.

\bibliographystyle{apsrev4-1}

\bibliography{shock}

\end{document}